# Fabrication and Room-Temperature Single-Charging Behavior of Self-Aligned Single-Dot Memory Devices

*Xiaohui Tang[1], Nicolas Reckinger[1], Vincent Bayot[1], Christophe Krzeminski[2], Emmanuel Dubois[2], Alexandre Villaret [3]  and Daniel-Camille Bensahel [3]*

[1]Microelectronics Laboratory (DICE) and CERMIN, Université Catholique de Louvain,
Place du Levant 3, 1348 Louvain-la-Neuve, Belgium
Tel: +32(0)10472555; Fax: +32(0)10472598
E-mail address: Tang@dice.ucl.ac.be

[2]Institut Supérieur d'Electronique du Nord, France

[3]ST Microelectronics, France

**Abstract:**

Self-aligned single-dot memory devices and arrays were fabricated based on arsenic-assisted etching and oxidation effects. The resulting device has a floating gate of about 5-10 nm, presenting single-electron memory operation at room temperature. In order to realize the final single-electron memory circuit, this paper investigates process repeatability, device uniformity in single-dot memory arrays, device scalability, and process transferability to an industrial application.

**Key words:** nanotechnology, single-electron memory, Coulomb blockade effect, self-aligned floating gate, arsenic-assisted etching and oxidation effects.

# 1. Introduction

Single-electron memories [1] are the ultimate device for low power operation. In single-electron memories, a nanometer-scale floating gate is charged/discharged by a few (ideally one) electrons based on energy quantization and Coulomb blockade effect [2]. Two main approaches, single-dot and multiple-dot devices, have been proposed to fabricate these devices since the first single-electron memory was reported in 1993 [3].

In multiple-dot devices [4]-[11], randomly deposited nanocrystals store individual electrons. A large number of dots implies that these devices are intrinsically not single-electron devices, but each dot stores only one or a few electrons depending on its size. Multiple-dot devices present the advantage of being more immune to spurious discharging as electrons are distributed among isolated dots. However, the randomness of nanocrystal deposition induces device characteristic fluctuations.

Single-dot devices are composed of a narrow channel MOSFET with a single nano-floating gate embedded in the gate dielectric. The key technological issue is to fabricate this nano-floating gate and to align it with the channel in a controlled manner. To our knowledge, contrary to multiple-dot memories, there are only a few papers related to Si single-dot single-electron memory devices [12]-[17], probably because of the difficulty of their realization. In the race for the ultimate memory device, single-dot memory devices have the advantages of nanometer dimension, stable performance and easy analysis over multiple-dot memories [1].

In most cases [12]-[16], very high-resolution lithography tool is required to define the floating gate and the channel. By making use of arsenic-assisted etching and oxidation effects, we have developed the fabrication process of a single-dot memory device [17]. It can be realized by current optical lithography tools, moreover, the floating gate is self-aligned with the channel. To realize the final single-electron memory circuit, we investigate here process repeatability, device uniformity in single-dot memory arrays, device scalability, and finally process transferability to an industrial production line. We also characterize miniaturized single-electron memory devices.

## 2. Device process

*A. Fabrication*

A detailed process can be found in [17]. The layout of the processed memory device is shown in the inset of Fig. 1.a. Wider source and drain regions are connected by a narrow-wide-narrow channel. The memory devices and arrays were fabricated on As-doped SOI wafer with a silicon overlay of 200 nm (at doping energy of 110 keV). The silicon overlay, covered by silicon nitride, is defined by a hybrid lithography (UV+e-beam) and reactive ion etching (RIE). Fig. 1.a shows a SEM cross-section of a device center after etching, which has a diabolo-like shape. Like all the cross-sectional SEM pictures in this study, contrast is enhanced by a quick HF dip that removes a bit of the silicon oxide surrounding Si parts. The arsenic-assisted etching effect leads to the fastest lateral etching rate of silicon where the arsenic is at its peak concentration (see the trench in Fig. 1.a). After wet oxidation, a tunnel oxide is formed in the trench region shown in Fig. 1.b. The tunnel oxide separates the top silicon dot from the bottom triangular wire. The dot and the triangular wire are used as the floating gate and the channel of the memory device, respectively. The floating gate is restricted to the central region since the top silicon dot is completely oxidized in the narrow regions of the channel (also called constrictions, see Fig. 3.b). In source and drain regions, the width of the silicon mesa is large enough to avoid the formation of the tunnel oxide at the location of arsenic peak concentration. Thus, the floating gate is separated from the channel in a single oxidation step. This results in a self-aligned structure.

*B. Repeatability*

To move the present process from laboratory concept to industrial application, it is necessary to assess the process repeatability. Therefore, we performed three experiments under the same conditions. Fig. 2 demonstrates a SEM image of the device center in the third experiment. Fig. 1.b was taken from the device in the second experiment. The first experimental result has been published in [17]. In these images, the same structure can be seen, indicating that this process is highly repeatable.

*C. Uniformity*

For the feasibility of a memory circuit, uniformity across the array is a more relevant parameter than the memory device size itself. Device uniformity is shown in Fig. 3.a and Fig. 3.b. In this experiment, different arrays were fabricated together with memory devices. Fig. 3.a exhibits the cross-section of the central part of the device for one array. It is worth noting that in the third line from left hand, the floating gate is nearly invisible. This comes from the limitations of our contrasting method (HF dip) which sometimes causes the removal of the very small silicon dot. However, this has no effect on real devices. Fig. 3.b illustrates the constriction region. As expected, only a triangular channel is formed since the silicon dot is completely consumed. It is found that the channel size is very uniform across each array.

Fig. 4 gives some typical geometrical parameters (i.e. the channel base width, the channel height and the tunnel oxide thickness) for 8 devices chosen randomly from a large number of devices. We cannot provide precise data for the silicon dot because the dot size is too small to be extracted accurately from SEM pictures. However, the equivalent diameter of the floating gate is estimated to be about 20 nm. In order to meet the requirements for room-temperature single-electron operation, the floating gate must be scaled down below 10 nm [[1]8].

*D. Scalability*

In the present process, the size of the floating gate depends on arsenic doping dose and energy, the thickness of the silicon overlay, as well as oxidation conditions. The device scalability is first investigated by simulation and then confirmed by experiments. The optimal scaling rule tells us that device miniaturization should be done by reducing silicon overlay thickness and implantation energy in the same ratio. By thinning the silicon overlay from 200 nm to 100 nm and decreasing the arsenic implantation energy from 110 keV to 55 keV, we fabricated 70 miniaturized devices. They present single-electron memory operation at room temperature. Hereafter, the device we discuss about refers to the miniaturized device.

The most critical step in this process is the wet oxidation since it governs the creation of the silicon dot on top of the channel. Oxidation is simulated by a commercial simulator (ISE DIOS) [[1]9]. Fig. 5 gives the simulation result and Fig. 6 shows a SEM image of the top view for one of the 70 fabricated devices before polysilicon gate deposition. The floating gate, which can barely be observed through the gate oxide, has a diameter around 10 nm with a pretty large uncertainty given the weak contrast between silicon and oxide. Even if the real dot size cannot be determined, the room-temperature single-electron behavior (see section 3) will confirm that this size is smaller than 10 nm. This dimension approaches to the limitation of the present process. To further reduce the size of the floating gate, a longer oxidation time will be needed. This would result in a significant increase in the tunnel oxide thickness, thereby, programming/erasing speed of the memory device would be decreased. Finally, Fig. 7 demonstrates a fabricated device with a control gate of 150 nm.

*E. Transferability*

The key steps of the present process, such as implantation, lithography, etching and oxidation, have also been tested in an industrial fabrication line of ST Microelectronics. Fig. 8 gives the SEM cross-section of a device center realized by current optical lithography tools. A floating gate of about 5 nm over the triangular channel can be seen clearly. Although further improvements are needed to optimize dot size and the tunnel oxide thickness, the present process is compatible with CMOS technology and can be considered as a potential option in the forthcoming technological nodes.

## 3. Electrical characterization

The condition to observe Coulomb blockade effect is $k_BT < q^2/C_\Sigma$. For real device applications, the charging energy $q^2/C_\Sigma$ should exceed the thermal energy $k_BT$ at least by a factor of three [18]. Here $C_\Sigma$ is the total capacitance of the floating gate. A first simple model for the floating gate is a silicon conducting sphere which has a self-capacitance of $C_\Sigma = 2\pi\varepsilon_{ox}d$ (d is the diameter of the floating gate and $\varepsilon_{ox}$ is the permittivity of oxide). For the presence of Coulomb blockade effect at room temperature, the diameter of the floating gate must be less than 8.5 nm from $k_BT = 0.3q^2/C_\Sigma$. This is consistent with the observed size of the dot (about 5-10 nm) from Fig. 6 and Fig. 8.

Fig. 9 shows the $I_d$-$V_g$ curve of a typical device at room temperature. The $I_d$-$V_g$ characteristics are that of a typical P-channel MOSFET (As-doped). Several jumps (see arrows) in the current, corresponding to single charge (hole) injection into the floating gate, can be seen in the figure. To further support single-charge injection, the time evolution of the drain current is carried out. In this measurement, a positive voltage of 5 V is first applied to the control gate for returning holes to the channel and obtaining an empty floating gate. Then a constant voltage of –5.58 V is set and held on during the measurement. Fig. 10 demonstrates a typical $I_d$-t characteristics. When one hole is injected into the floating gate, an abrupt reduction of drain current is observed. The quantization of the current steps, of $\Delta I_d \cong 0.94$ nA, confirms that they stem from single hole injection in the nano-floating gate.

From the $I_d$-t measurements, we found that the hole injection is stochastic. By repeating the same measurement more than 100 times, we calculated the mean injection time $\tau$ of the first hole to be 0.7 s at a gate voltage of –5.58 V, which is much smaller than Yano's value (12.9 s) at a gate voltage of 12 V [**2**0]. It is possible to model hole injection process by a simple Poisson process. To calculate the probability of the first hole injection in the floating gate ($P_1$), we simply use $P_1(t) = 1 - P_0(t)$, where $P_0(t)$ is the probability for an empty floating gate at t time ($P_0(t) = \exp(-t/\tau)$). The Poisson law (the solid line) and the results extracted from the experiment (the star symbols) are shown in Fig. 11. As we can see in the figure, the experimental data are in good agreement with this approximation before 1 s, while the data fitting becomes poor after 1 s since we have neglected the probabilities for the second, third, … hole injection in the calculation. This indicates that the other hole injection except for the first hole should be taken into account after 1 s.

The $I_d$-t is also measured at different gate voltages. The extracted values of drain current steps (a bunch of current values) as a function of gate voltage are shown in Fig. 12. Note that it is impossible to extract all current values at some gate voltages due to fast injection. The solid lines, plotted by linking the current values from each hole injection status, represent the $|I_d|$-$V_g$ curves with zero, one, two, three … holes in the floating gate, respectively.

For an SOI standard transistor, which has a gate length of 1 µm, the threshold voltage is determined as the gate voltage at $I_d = 100$ nA/µm and $V_d = 50$ mV [**2**1]. Taking into account the specificities of our device structure, we set the threshold voltage as the gate voltage at $I_d = 1$ nA and $V_d = 10$ mV. When the floating gate is empty, the threshold voltage is –4.95 V, while when the first hole is injected into the floating gate, the threshold voltage becomes –5.05 V. For small $|V_g|$, the threshold voltage shift is found to be approximately equal to n·$\Delta V_n$, where n is the number of injected hole and $|\Delta V_n| \cong 0.1$ V (for n = 1,2,3). The threshold voltage shift is basically proportional to the number of holes injected, which is in qualitative agreement with the theoretical calculation result in [2]. This may be explained by the special hole distribution in the triangular channel surrounded by an $\Omega$-gate. In this device, the holes of the initial inversion layer induced by the gate voltage occur mainly at the tip of the triangular channel. Hence, the channel holes experience the same screening for every hole injected in the floating gate. However, for large $|V_g|$, the proportional relation (n·$\Delta V_n$) is untenable. As seen in the figure, the fourth voltage shift $|\Delta V_4|$ is larger than 0.1V. This issue is under investigation.

The hysteretic behavior of the device at room temperature is shown in Fig. 13. When $V_g$ is swept from –3V to –7V, five $I_d$ shifts are observed, while two back shifts are observed when $V_g$ sweep is reversed. Hysteresis has been observed on almost all devices, while the reference device without silicon dot does not exhibit any hysteresis. This indicates that holes are injected into the floating gate rather than into possible traps of the gate oxide. Finally, it is worth noting that the threshold voltage shift and the hole charging voltage are varying a little from one device to another. To reach reproducibility of electrical characteristics, the layout of the device should be transferred perfectly to the silicon overlay by using a good lithography tool, which was not possible with our lithography tool. The relatively short retention time, compared to our results on the 20 nm-dot devices [17], is probably due to (1) a poor quality of tunnel oxide, because wet oxidation at low temperature (800°C) was used to form the tunnel oxide; (2) a high arsenic doping concentration inducing stress, point-defects and dislocations. However, the tunnel oxide property can be improved by increasing oxidation temperature and adjusted by tuning oxidation time.

## 4. Conclusion

Self-aligned single-dot memory devices and arrays with a floating gate of about 5-10 nm have been fabricated. The quantized threshold voltage shift and the hysteresis characteristics imply that room-temperature single-hole memory operation is achieved. Process repeatability and size uniformity across each array indicate that this process can be used to realize memory circuits. The present process is compatible with CMOS technology and several key steps have been tested in an industrial fabrication line.

## ACKNOWLEDGMENTS

The authors would like to acknowledge A. Crahay, D. Spote, B. Katschmarskyj, P. Loumaye, C. Renaux, N. Mahieu and M. Zitout  for their help in the device fabrication.

## Captions

Fig. 1: SEM cross-section of a device centre: (a) after etching, the trench is observed where the As peak concentration is located; (b) after oxidation, the trench is completely consumed, forming the tunnel oxide.

Fig. 2: SEM cross-section of a device centre after oxidation. The floating gate and the channel are clearly separated by the tunnel oxide.

Fig. 3: SEM cross-sections of lines within an array: (a) with the same width as the device centre and (b) with the same width as the constrictions of the device. In both cases, the device structure is uniformly repeated.

Fig. 4: Typical geometrical parameters for 8 devices realized in different regions of one wafer. The data show small variations from one device to another due to the limit of lithography system.

Fig. 5: Simulated cross-section for one device with a silicon overlay of 100 nm.

Fig. 6: SEM top view of a device before polysilicon gate deposition. The floating gate can be seen through the gate oxide in the centre of the memory device. The dot size is approximately equal to 10 nm.

Fig. 7: SEM top view of a fabricated memory device, the control gate has a length of 150 nm.

Fig. 8: SEM cross-section of a device centre processed by ST Microelectronics. The floating gate size is about 5 nm.

Fig. 9: $I_d$-$V_g$ characteristics (solid line) of a typical device at room temperature. The arrows represent one, two, three ... holes injected into the floating gate.

Fig. 10: Time evolution of drain current for a constant gate voltage of –5.58V. The different steps in $I_d$-t are an evidence for single hole injection in the floating gate.

Fig. 11: First hole injection probability, accounted by a simple Poisson process. A mean injection time of 0.7 s, obtained from more than 100 $I_d$-t measurements, is used in the calculation. The solid line shows the calculation result and the star symbols represent data extracted from the experiment.

Fig. 12: Extracted values of drain current steps as a function of gate voltage at room temperature. The solid lines represent the $|I_d|$-$V_g$ curve with zero, one, two, three and four holes in the floating gate, respectively.

Fig. 13: Hysteresis characteristics of a device at room temperature at $V_d$ = -10 mV. Many $I_d$ shifts can be seen when the gate voltage is swept forth and back.

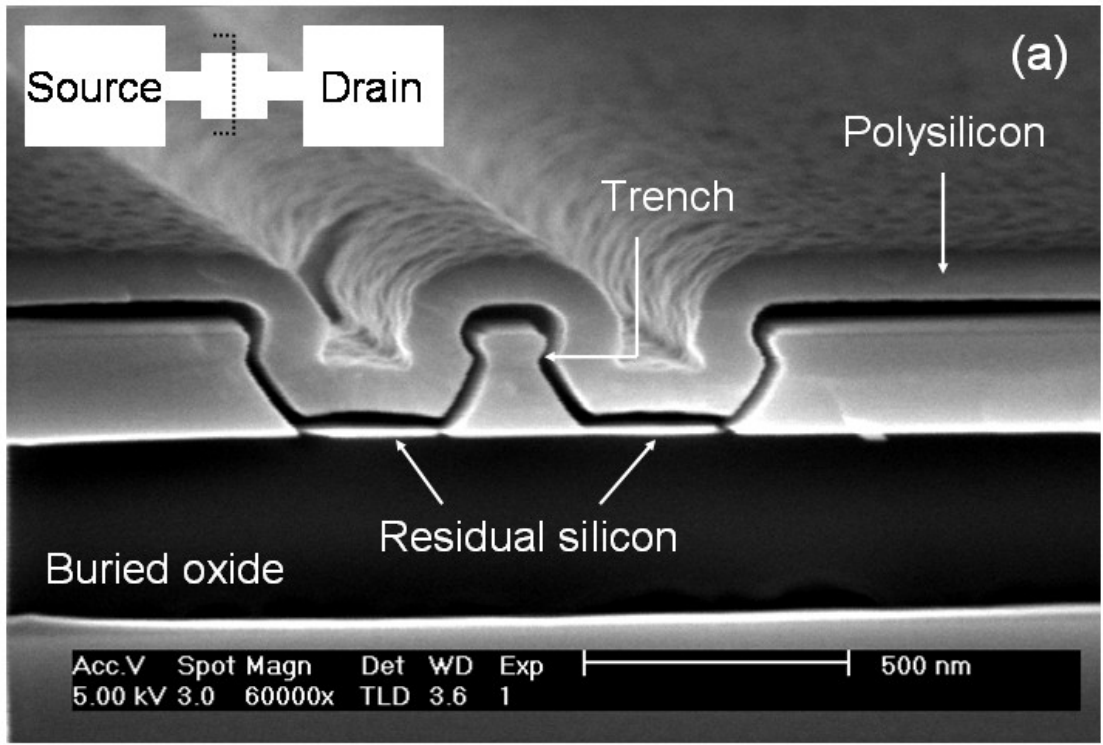

Fig. 1.a : Xiaohui Tang et al.

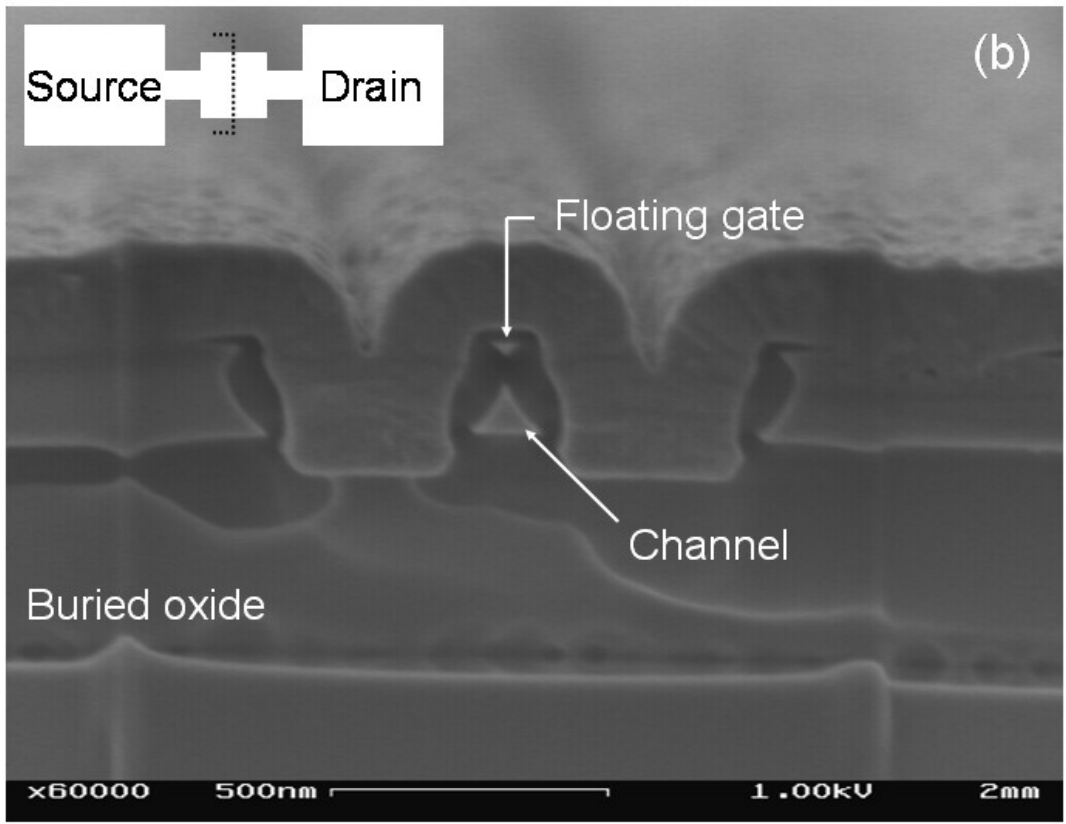

Fig. 1.b : Xiaohui Tang et al.

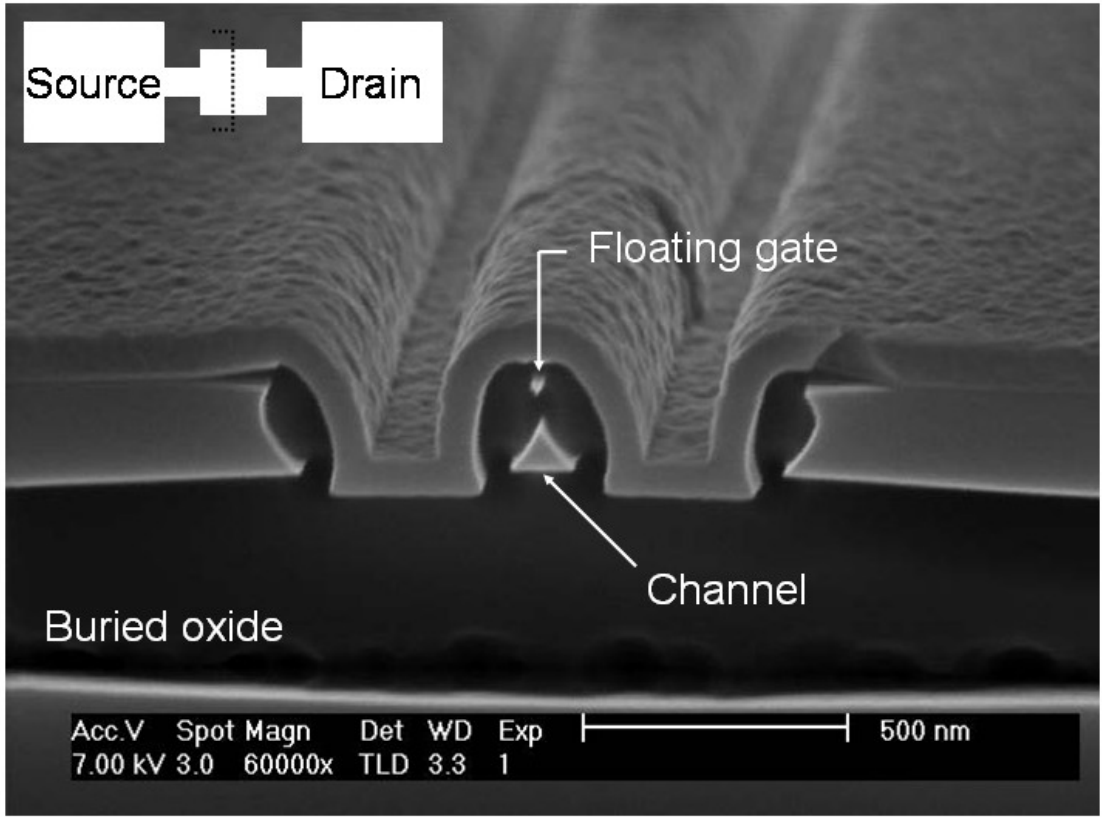

Fig. 2 : Xiaohui Tang et al.

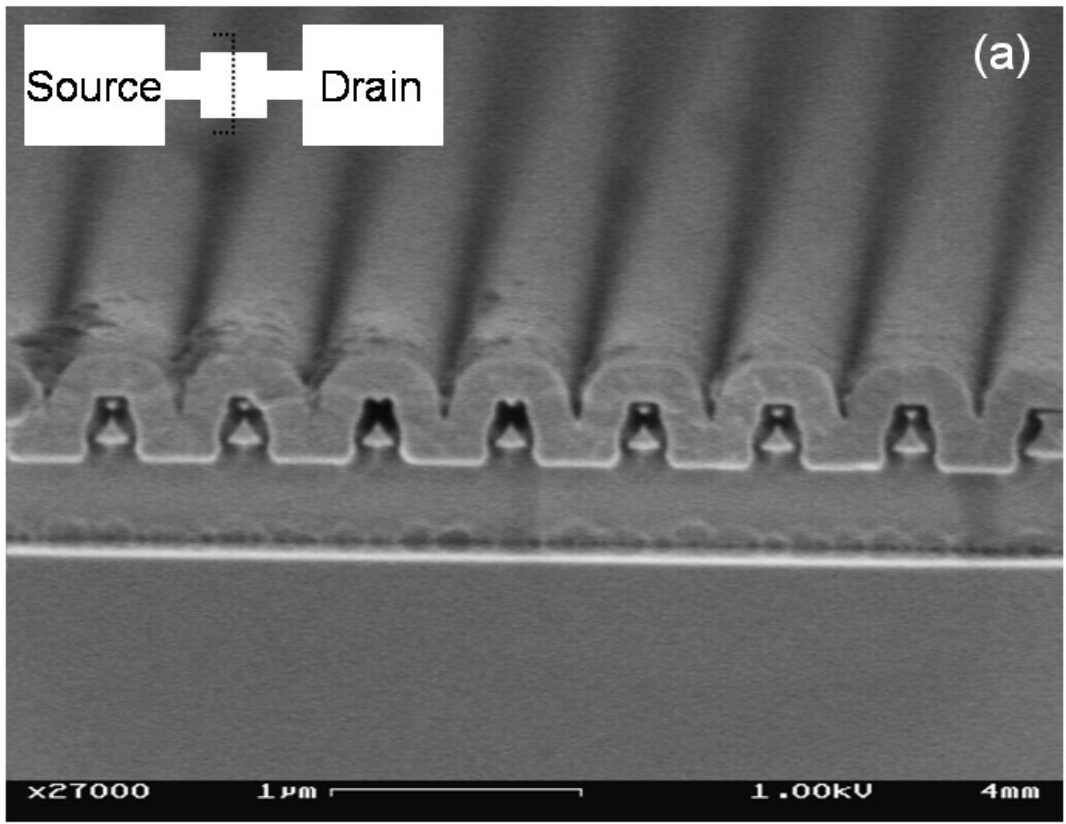

Fig. 3.a : Xiaohui Tang et al.

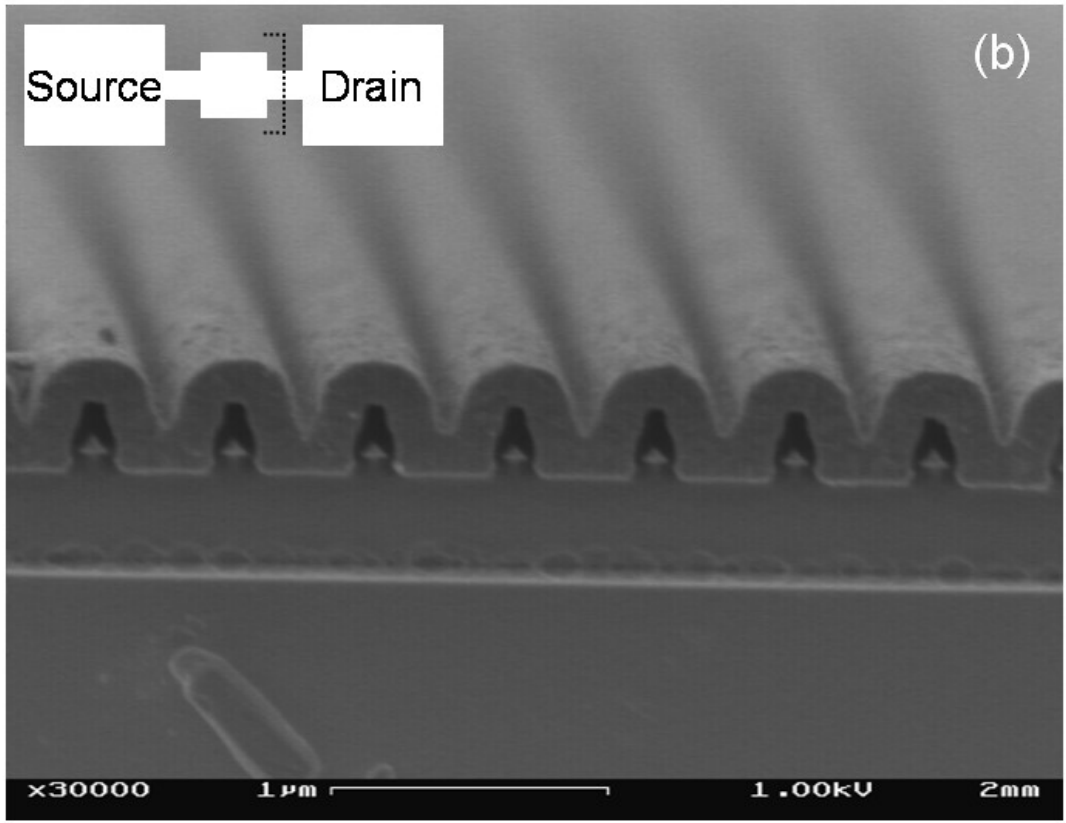

Fig. 3.b : Xiaohui Tang et al.

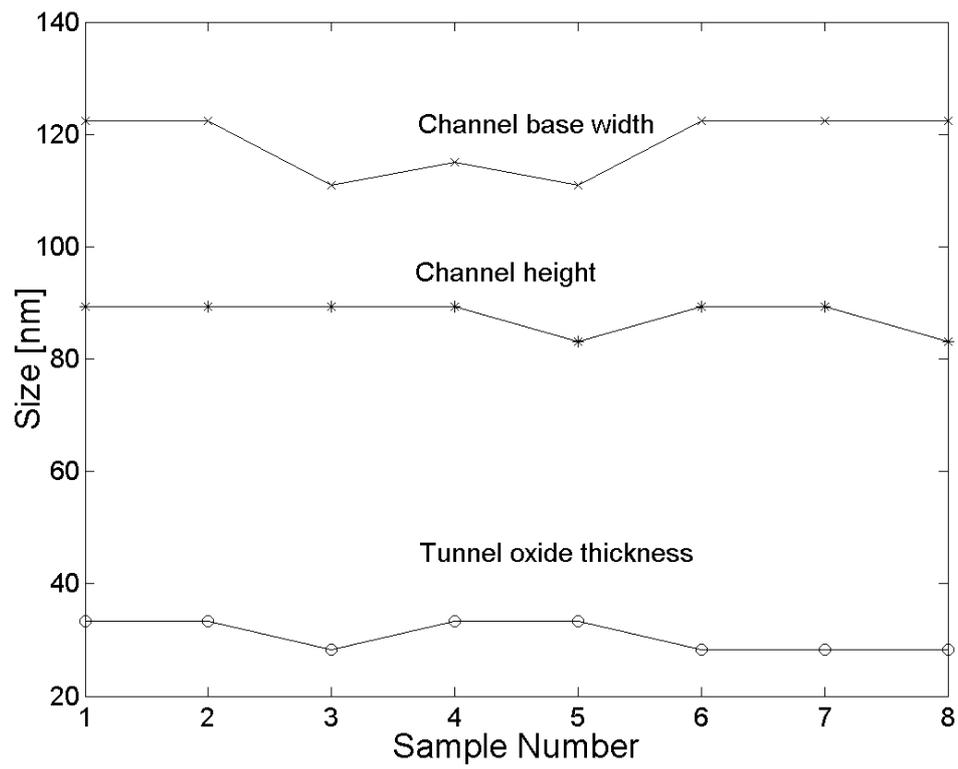

Fig. 4 : Xiaohui Tang et al.

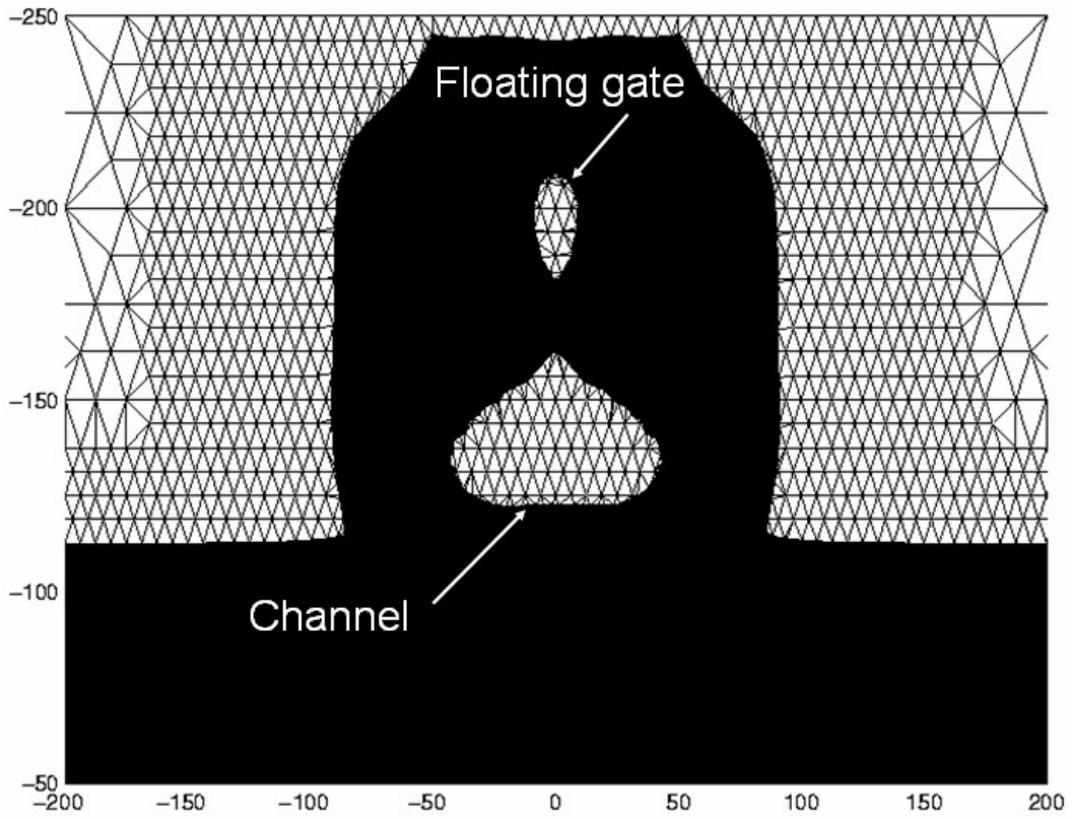

Fig. 5 : Xiaohui Tang et al.

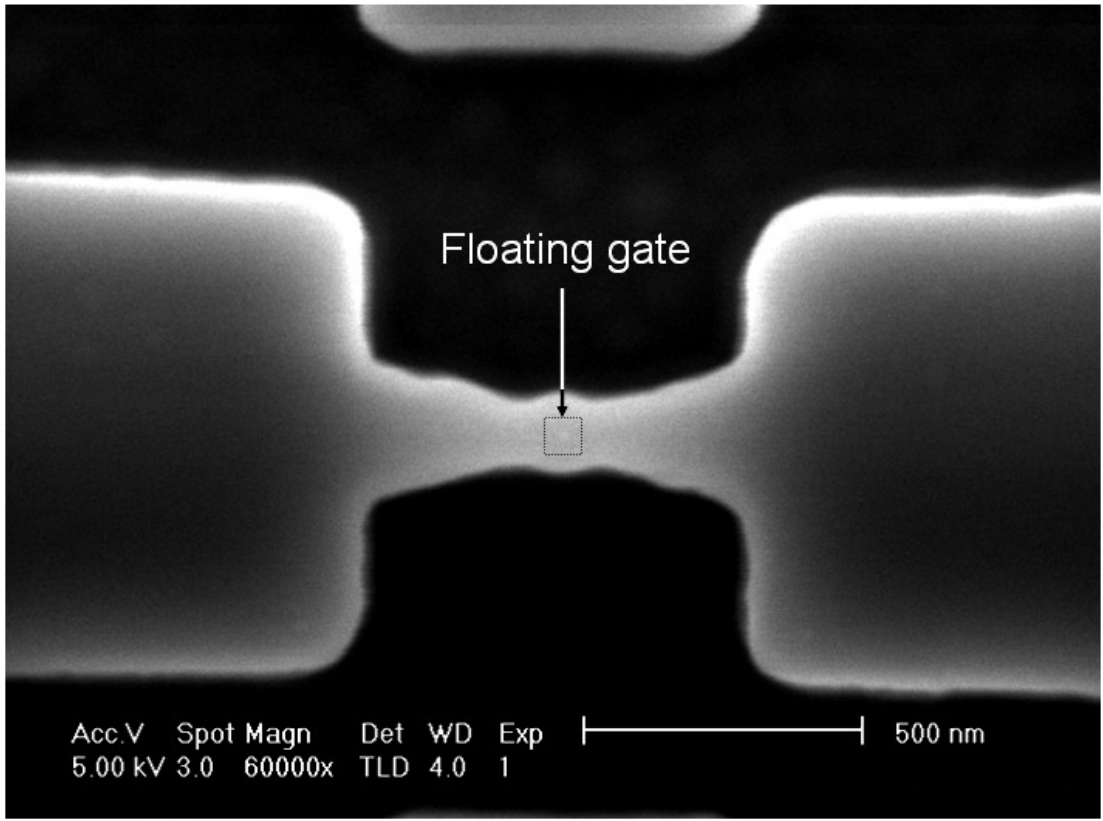

Fig. 6 : Xiaohui Tang et al.

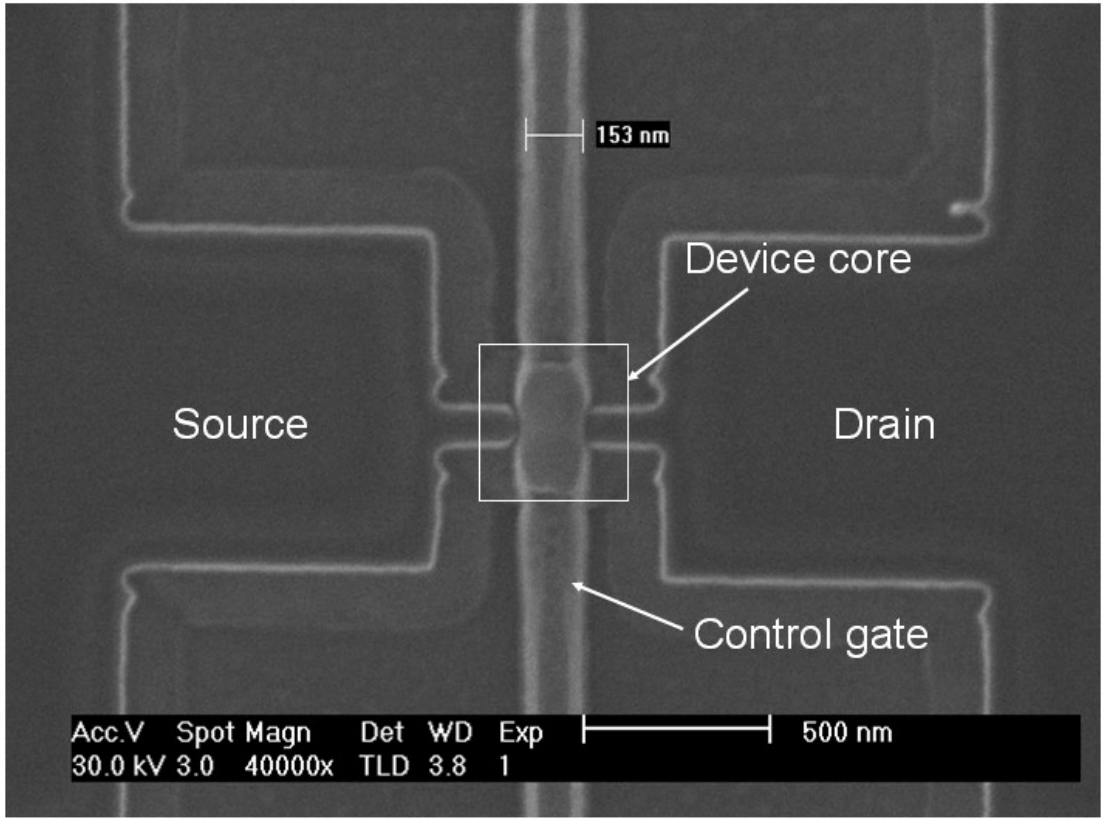

Fig. 7 : Xiaohui Tang et al.

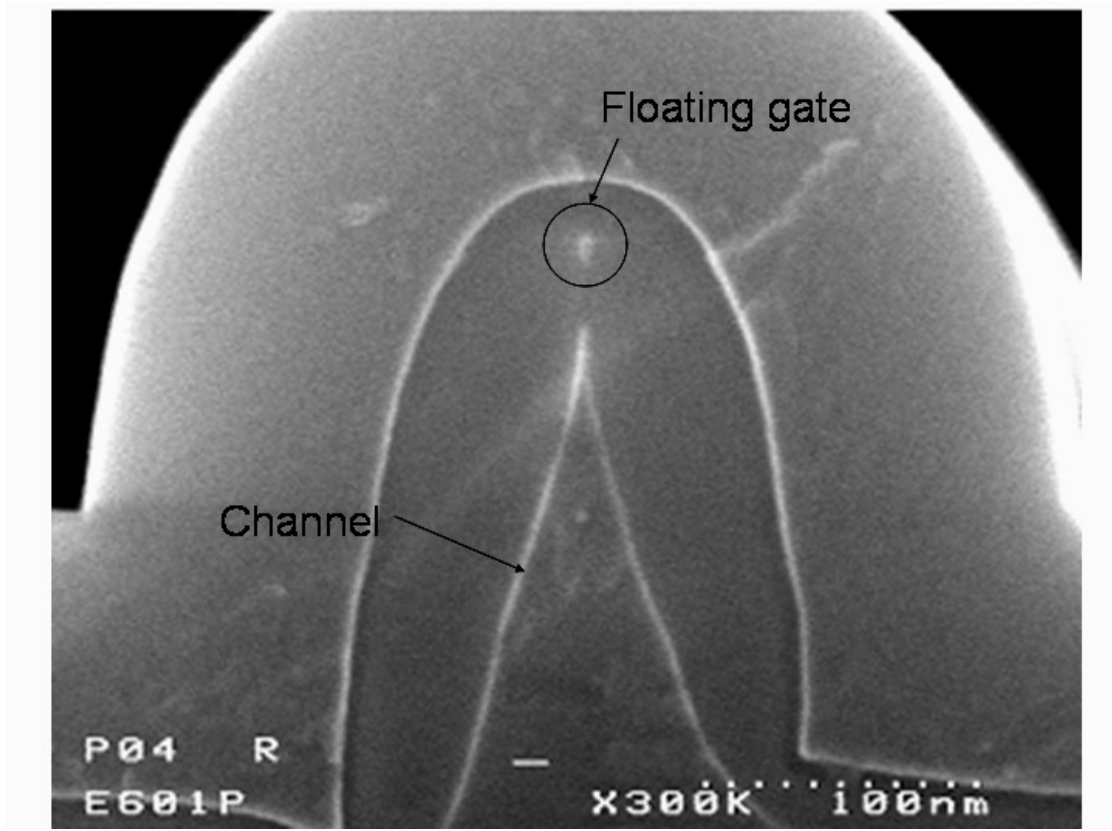

Fig. 8 : Xiaohui Tang et al.

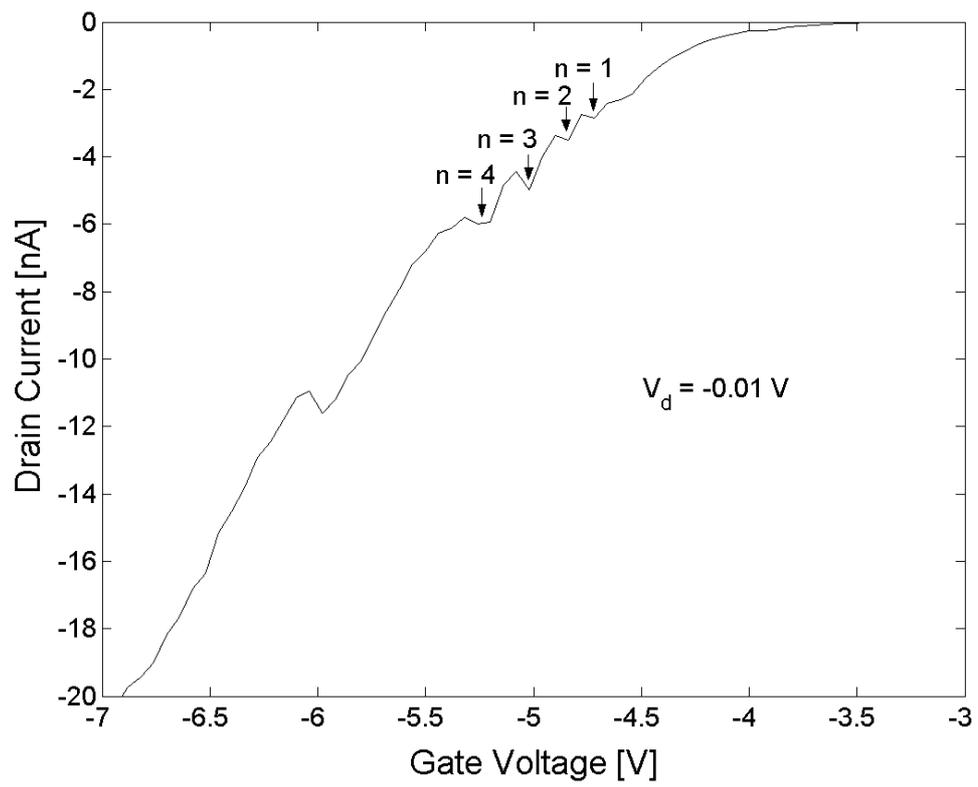

Fig. 9 : Xiaohui Tang et al.

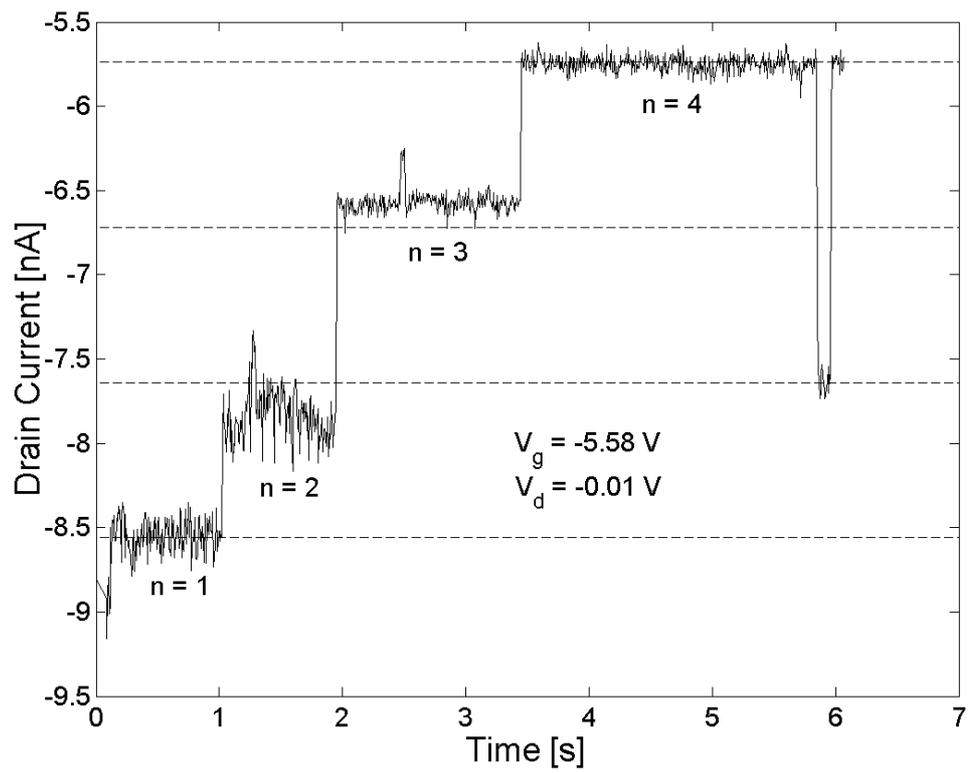

Fig. 10 : Xiaohui Tang et al.

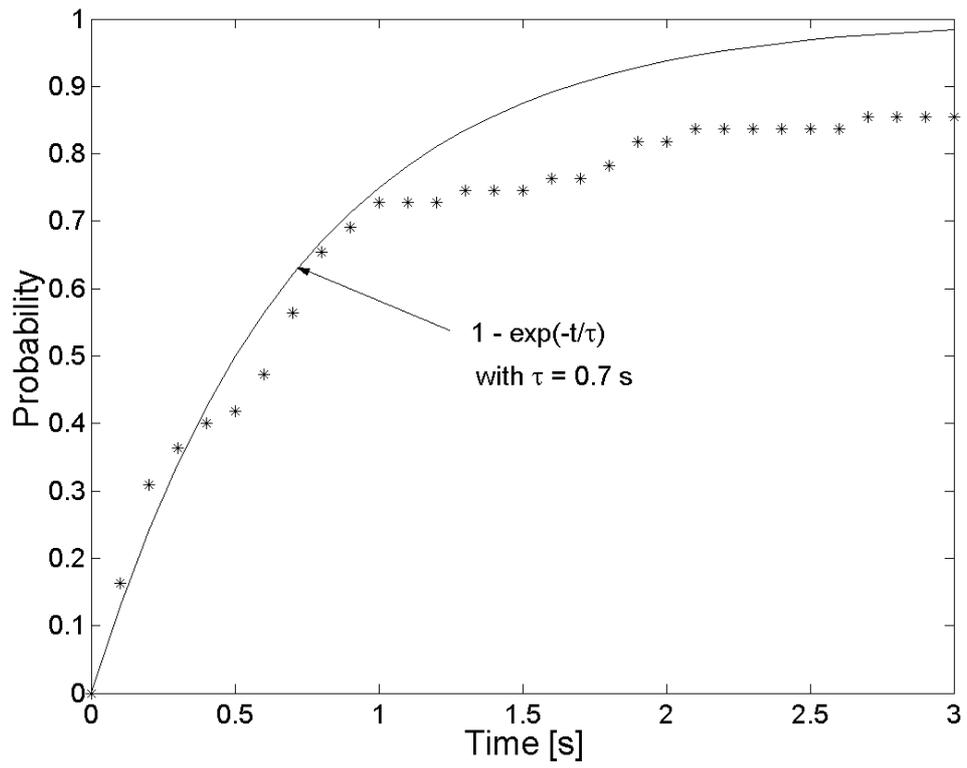

Fig. 11 : Xiaohui Tang et al.

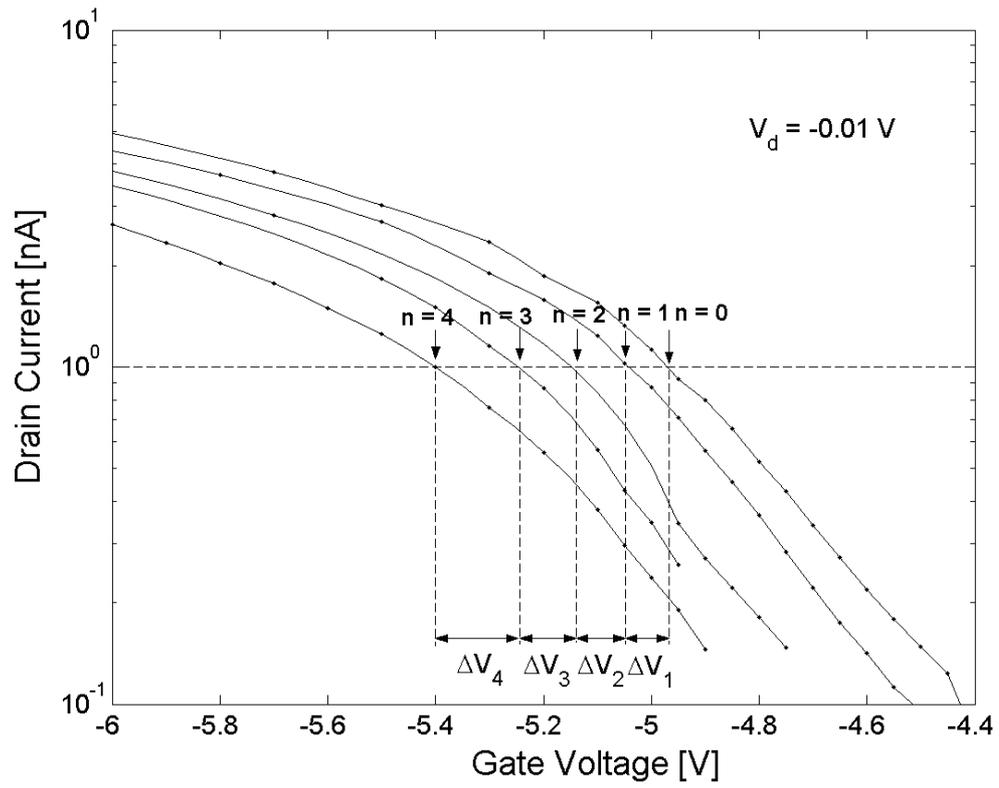

Fig. 12 : Xiaohui Tang et al.

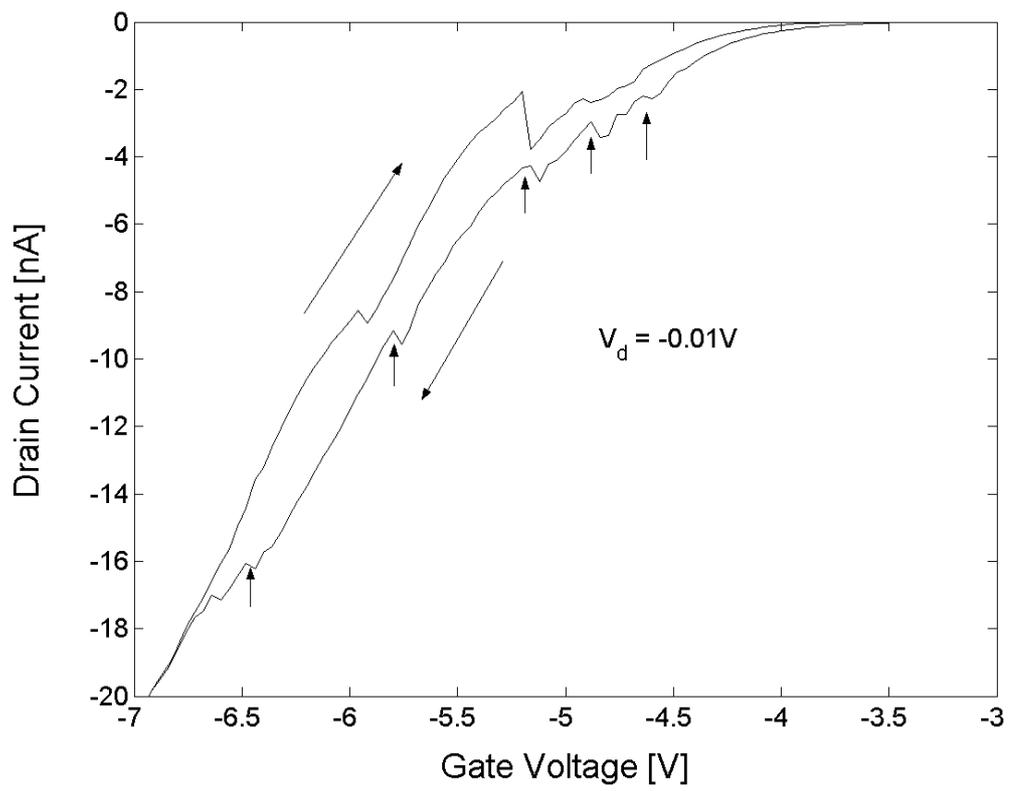

Fig. 13 : Xiaohui Tang et al.